\documentclass[a4paper,11pt]{article}
\usepackage{pos}
\usepackage{caption}
\usepackage{subcaption}

\newcommand{\RN}[1]{\uppercase\expandafter{\romannumeral#1}}

\title{Computing hybrid static potentials at short quark-antiquark separations from fine lattices in $SU(3)$ Yang-Mills theory}
 \ShortTitle{Hybrid static potentials at short quark-antiquark separations}

\author*[a,b]{Carolin Schlosser}
\author[a,b]{Marc Wagner}

\affiliation[a]{Institut für Theoretische Physik, Goethe-Universität Frankfurt am Main,\\
  Max-von-Laue-Straße 1, D-60438
  Frankfurt am Main, Germany}

\affiliation[b]{Helmholtz Research Academy Hesse for FAIR, Campus Riedberg, \\
	 Max-von-Laue-Straße 12, D-60438 
	 Frankfurt am Main, Germany}

\emailAdd{schlosser@itp.uni-frankfurt.de}
\emailAdd{mwagner@itp.uni-frankfurt.de}

\abstract{
	We compute hybrid static potentials in $SU(3)$ lattice Yang-Mills theory at short quark-antiquark separations using four different small lattice spacings as small as $0.04\,\text{fm}$. 
	The resulting static potentials are important, e.g.\ when studying heavy hybrid mesons in the Born-Oppenheimer approximation. 
	We also discuss and exclude possible systematic errors from topological freezing, the finite lattice volume and glueball decays.
	}

\FullConference{%
 The 38th International Symposium on Lattice Field Theory, LATTICE2021
  26th-30th July, 2021
  Zoom/Gather@Massachusetts Institute of Technology
}

\begin{document}
\maketitle

\section{Introduction}
	Hybrid static potentials represent the energy of an excited gluon field in the presence of a static quark and a static antiquark as a function of their separation and are, thus, closely related to heavy hybrid mesons.
	The gluonic excitation contributes to the quantum numbers of the meson such that also exotic combinations of $J^{PC}$ are possible, which can not be studied with the constituent quark model.
	Exotic mesons as well as gluonic excitations are currently an active field of experimental and theoretical investigation (for reviews cf.\ e.g.~\cite{Braaten:2014ita,Meyer:2015eta,Swanson:2015wgq,Lebed:2016hpi,Olsen:2017bmm,Brambilla:2019esw}).
	
	The masses of heavy hybrid mesons with heavy bottom and charm quarks can be computed e.g.\ in the Born-Oppenheimer approximation~\cite{PERANTONIS1990854,Juge:1997nc,Juge:1999ie,Braaten:2014qka,Capitani:2018rox}.
	In a first step,
	the heavy quarks are considered as static and the energy of the gluonic field can be obtained with lattice gauge theory.
	In a second step, the radial Schrödinger equation for the relative coordinate of the heavy quark-antiquark pair is solved with a parametrization of the lattice data for the corresponding hybrid potential.
	The resulting energy eigenvalues are directly related to masses of heavy hybrid mesons.
	
	Moreover, lattice field theory results for hybrid potentials are important as input for effective field theories, where the lattice results at short distances are required to fix matching coefficients of potential Non-Relativistic QCD (pNRQCD)~\cite{Berwein:2015vca,Oncala:2017hop,Brambilla:2018pyn,Brambilla:2019jfi}.

	In this work, we compute hybrid static potentials in $SU(3)$ lattice gauge theory at four different lattice spacings $a= 0.040 \, \text{fm}$, $0.048 \, \text{fm}$, $0.060\,\text{fm}$ and $0.093\,\text{fm}$, which include significantly smaller lattice spacings than used in previous works, e.g.\ in Refs.~\cite{Perantonis1989StaticPF,PERANTONIS1990854,Juge:1997nc,Juge:1997ir,Morningstar:1999rf,Juge:2002br,Bali:2003jq,Juge:2003ge,Capitani:2018rox}.
	We present lattice field theory results for the ordinary as well as the two lowest hybrid static potentials at quark-antiquark separations as small as $r\approx 0.08\,\text{fm}$ and discuss and exclude possible sources of systematic errors such as effects from the finite lattice size, topology freezing and glueball decays.

\section{Hybrid static potentials}
	Quantum numbers of (hybrid) static potentials are the following:
	\begin{itemize}
		\item $\Lambda = \Sigma (=0), \Pi (=1), \Delta (=2), \dots $ denotes non-negative integer values of the total angular momentum with respect to the quark-antiquark separation axis. 
		\item $\eta= g,u$ describes the even ($g$) or odd ($u$) behavior under the combined parity transformation and charge conjugation, $\mathcal{P}\circ \mathcal{C}$.
		\item $\epsilon=+,-$ is the eigenvalue of a reflection $\mathcal{P}_x$ along an axis perpendicular to the quark-antiquark separation axis. 
		Hybrid static potentials with $\Lambda \ge 1$ are degenerate with respect to $\epsilon$.
	\end{itemize}
	The ordinary static potential is labeled by $\Lambda_{\eta}^{\epsilon}=\Sigma_g^+$, while the two lowest hybrid static potentials have quantum numbers $\Pi_u$ and $\Sigma_u^-$.
	
	Hybrid static potentials are computed from Wilson loop-like correlation functions with non-trivial spatial transporters, i.e.\ different from a straight line.
	In particular, to excite gluons with quantum numbers $\Pi_u$ and $\Sigma_u^-$, we employ the creation operators $S_{\text{\RN{3}},1}$ and $S_{\text{\RN{4}},2}$, respectively, which are defined in Table $3$ and Table $5$ of Ref.~\cite{Capitani:2018rox}, where the creation operators are discussed in detail.
	
\section{Numerical results}
	\subsection{Lattice setup}
	All computations were performed on $SU(3)$ gauge link configurations, which were generated with a Monte Carlo heatbath algorithm with the standard Wilson plaquette action using the CL2QCD software package~\cite{Philipsen:2014mra}.
	In addition, we used the multilevel algorithm \cite{Luscher:2001up} to achieve an error reduction in the correlation functions.
	We generated four ensembles with gauge couplings $\beta=6.000$, $6.284$, $6.451$ and $6.594$, which correspond to lattice spacings $a \approx 0.093\,\text{fm},\ 0.060\,\text{fm},\ 0.048\,\text{fm}$ and $0.040\,\text{fm}$, respectively, when using a parametrization of $\ln(a/r_0)$ from Ref.~\cite{Necco:2001xg} and setting $r_0=0.5\,\text{fm}$.
	The physical lattice volume is the same for each ensemble, i.e.\ $L^3 \times T \approx (1.2\,\text{fm})^3 \times (2.4\,\text{fm})$.
	
	The (hybrid) static potentials are obtained from plateaus of the corresponding effective potentials. 
	Contaminations by excited states were minimized by the use of optimized creation operators from Ref.~\cite{Capitani:2018rox}, where the optimization was performed at a lattice spacing equal to our coarsest lattice spacing, $a=0.093\,\text{fm}$. For the finer lattice spacings we adjust the spatial extent in lattice units to keep the physical size constant.
	Moreover, we applied an optimized number of APE-smearing steps~(see e.g.\ \cite{,Jansen:2008si}) with $\alpha_{\text{APE}}=0.5$ on the spatial gauge links, i.e.\ $N_{\text{APE}}$ was increased with decreasing lattice spacing.
	
	Furthermore, we define (hybrid) static potentials as functions of tree-level improved separations, which reduces lattice discretization effects particularly at small $r/a$~\cite{Necco:2001xg,Necco:2003jf}.

	\subsection{Hybrid static potentials}
\begin{figure}
		\includegraphics[width=\linewidth]{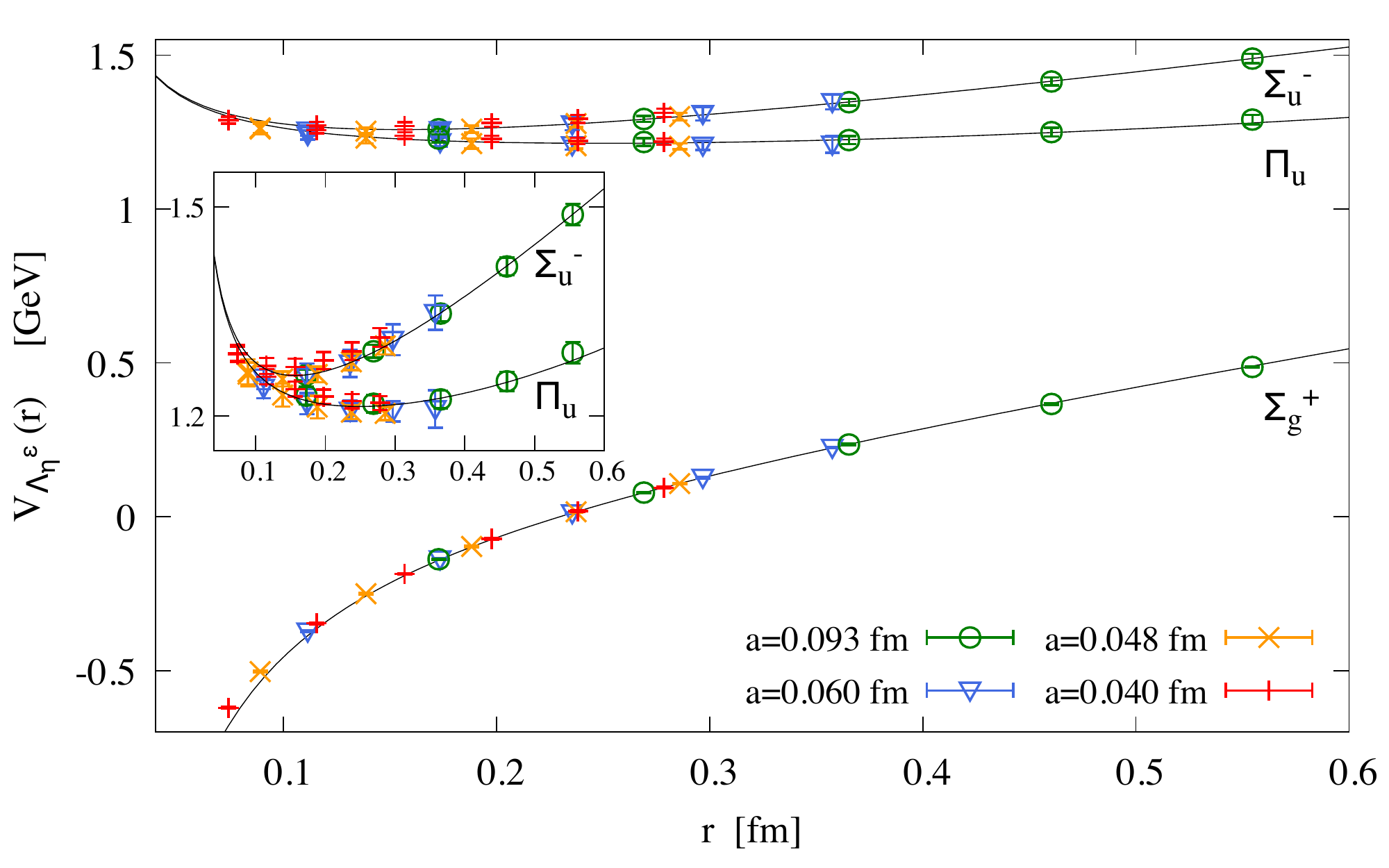}
		\caption{Lattice results for the (hybrid) static potentials.}
		\label{fig:potentials}
	\end{figure}

	In Figure \ref{fig:potentials}, we present our lattice results for the ordinary static potential $\Sigma_g^+$ and the two lowest hybrid static potentials $\Pi_u$ and $\Sigma_u^-$ for all four different lattice spacings as functions of the quark-antiquark separation $r$.
	We show static potentials at separations as small as $0.08\,\text{fm}$, which corresponds to twice the smallest lattice spacing, as discretization effects are expected to be non-negligible for $r < 2a$.
	The lattice results for the hybrid static potentials from our fine lattice spacings show the $1/r$-upward curvature as it is predicted by weakly-coupled pNRQCD at small $r$ \cite{Brambilla:1999xf,Berwein:2015vca}.
	Our results also indicate the degeneracy in the short distance limit, as it is expected for the two lowest hybrid static potentials.
	
	Parametrizations of our lattice results for the static potentials will be discussed in detail in a future publication.
	There we also use the parametrizations for the computation of heavy hybrid meson masses in the Born-Oppenheimer approximation.

\section{Excluding possible systematic errors}
	\subsection{\label{sec:finitevolume}Finite spatial lattice volume}
	When performing computations at different spatial lattice volumes for gauge group $SU(2)$ (see Ref.~\cite{Riehl:2020crt}) we observed a sizable volume dependence of the ordinary static potential and hybrid static potentials, when the volume is smaller than $\approx  (1.0 \,\text{fm})^3$.
	For example, shrinking the spatial lattice volume below this size causes a small negative shift for the ordinary static potential $\Sigma_g^+$, while there is a much larger positive shift for the hybrid static potential $\Pi_u$.
	For a spatial lattice volume of $(1.2 \,\text{fm})^3$, however, as used for the computation of the potentials shown in Figure~\ref{fig:potentials}, these finite volume corrections are tiny and negligible compared to statistical errors.
	
	\subsection{Topological freezing}
	Gauge field configurations can be classified according to their topological charge.
	\textit{Topological freezing} denotes the problem that a Monte Carlo simulation is trapped in one of the topological sectors.
	The gauge link configurations generated in such a simulation do not form a representative set distributed according to $e^{-S}$.
	This problem is typically observed when using a lattice spacing $a$ smaller than $\approx 0.05 \, \text{fm}$ \cite{Luscher:2011kk} and becomes more severe, when approaching the continuum limit, i.e.\ when further decreasing $a$.
	If a simulation is fully trapped in a topological sector, observables exhibit specific finite volume corrections in addition to those discussed in section~\ref{sec:finitevolume} (see e.g.\ Refs.~\cite{Brower:2003yx,Aoki:2007ka,Dromard:2014ela,Bietenholz:2016ymo}).
	
\begin{figure}\centering
		\includegraphics[width=0.6\linewidth]{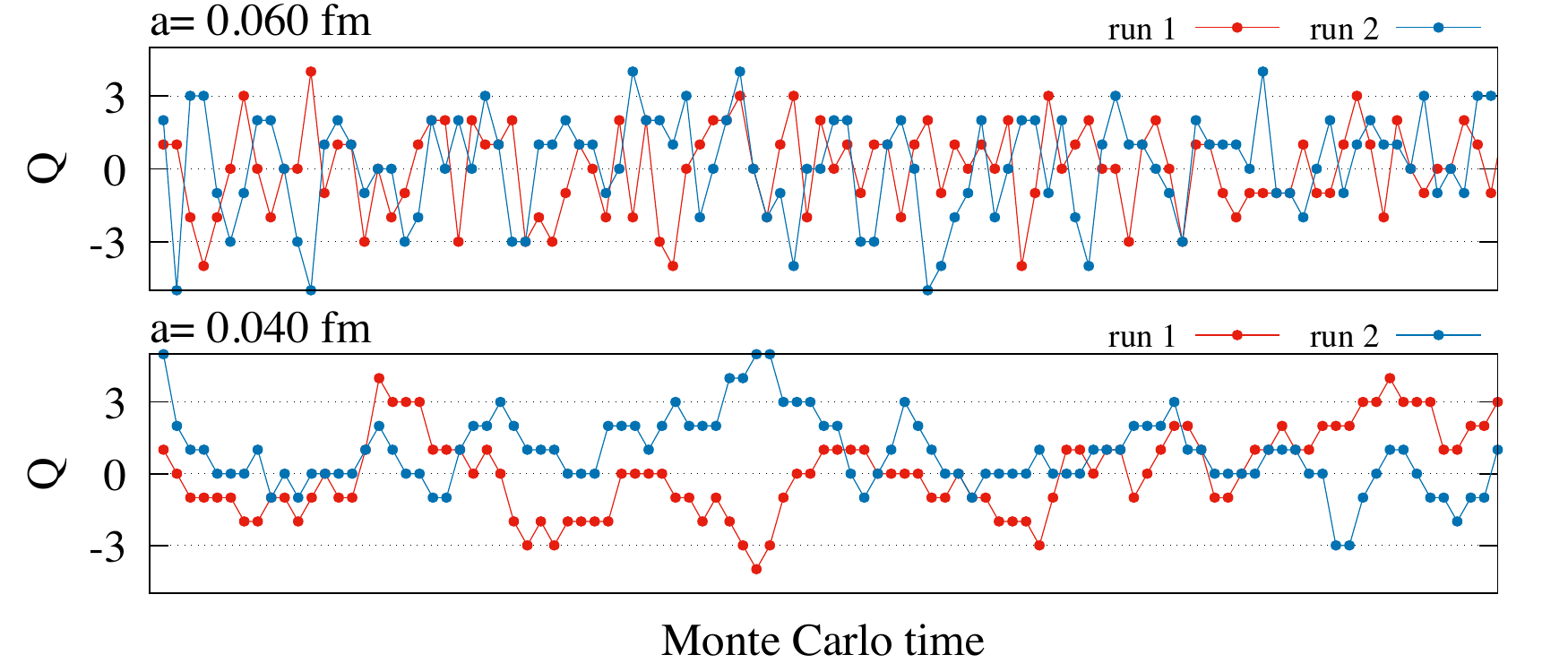}
		\caption{Topological charge as a function of the Monte Carlo time for two independent exemplary runs for two lattice spacings.}
		\label{fig:topcharge}
	\end{figure}

	To check, whether our simulations suffer from topological freezing, we computed the topological charge on the gauge link configuration via a field-theoretic definition with a simple clover-leaf discretization and 4-dimensional APE-smearing \cite{Cichy:2014qta}. 
	In Figure \ref{fig:topcharge}, we show the topological charge as a function of the Monte Carlo time for two independent exemplary runs for two lattice spacings.
	One can see that the topological charge still changes frequently even at the smallest lattice spacing, $a=0.040\,\text{fm}$.
	The topological charge distribution and topological susceptibility also indicate that the Monte Carlo algorithm is able to sample the gauge link configurations correctly.
	Using a suitable binning and several independent Monte Carlo runs we exclude that statistical errors are underestimated because of autocorrelations, which are also expected to increase with decreasing lattice spacing.
	In summary, the potentials presented in this work should be free of any systematic errors from topological freezing.
	
	\subsection{Glueball decays}
\begin{figure}\centering
		\includegraphics[width=0.6\linewidth,page=1]{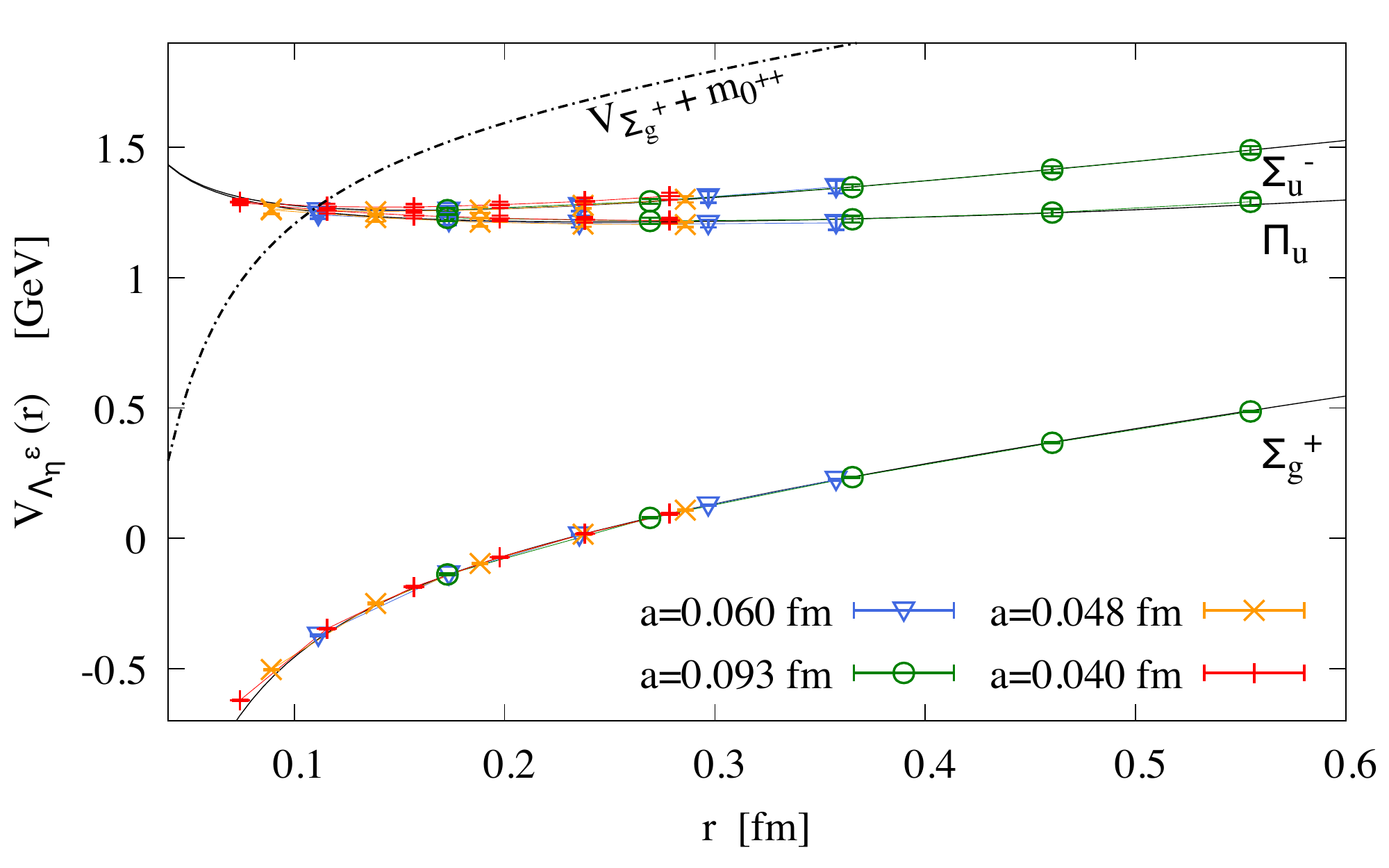}
		\captionof{figure}{Threshold energy $V_{\Sigma_g^+}(r) + m_{0^{++}}$ for the decay of a hybrid flux tube to the flux tube of the ordinary static potential and the lightest $0^{++}$ glueball. }
		\label{fig:glueballdecay}
	\end{figure}

	For sufficiently small quark-antiquark separations $r$ the energy difference between a hybrid static potential and the ordinary static potential $\Sigma_g^+$ is large enough such that the hybrid flux tube can dissolve into a glueball and the $\Sigma_g^+$ flux tube.
	The minimal energy, which is necessary for such a decay into the lightest glueball with quantum numbers $J^{PC}=0^{++}$ and mass
	$m_{0^{++}}\approx 4.21/r_0$~\cite{Morningstar:1999rf}
%	$m_{0^{++}}\approx 1.73\,\text{GeV}$~\cite{Morningstar:1999rf}
	is shown as a dashed line in Figure~\ref{fig:glueballdecay} together with our lattice results for hybrid static potentials.
	
	Below a critical separation, in particular below $r_{\text{crit}} \approx 0.1\,\text{fm}$ for the lowest hybrid potential $\Pi_u$, a $0^{++}$ glueball decay is energetically allowed and the hybrid static potential creation operator might generate non-vanishing overlap to the $\Sigma_g^+$ flux tube and the glueball.
	For the second lowest hybrid potential $\Sigma_u^-$ and the higher lying hybrid potential $\Sigma_g^-$ it is possible to exclude decays to the lightest $0^{++}$ glueball using symmetry arguments (we will discuss that in detail in a future publication). 
	Still allowed are decays into the next lightest glueball with quantum numbers $J^{PC}=2^{++}$. 
	However, this is energetically only possible at significantly smaller separations than those we investigated.
	
 	In this work, we present lattice results for the hybrid static potentials $\Pi_u$ and $\Sigma_u^-$ for separations below $r_{\text{crit}} \approx 0.1 \, \text{fm}$.
	There is, however, no sign of contamination by glueball decays, since the two lowest hybrid static potentials reveal the expected upward curvature and degeneracy at small separations \cite{Berwein:2015vca}.

	% ********************
	% ********************
	% ********************
	% ********************

\section*{Acknowledgments}
	We thank Christian Reisinger for providing his multilevel code and helpful conversations.
	
	M.W.\ acknowledges funding by the Heisenberg Programme of the Deutsche Forschungsgemeinschaft (DFG, German Research Foundation) -- Projektnummer 399217702.
	
	Calculations on the Goethe-HLR and on the FUCHS-CSC high-performance computer of the Frankfurt University were conducted for this research. We would like to thank HPC-Hessen, funded by the State Ministry of Higher Education, Research and the Arts, for programming advice.

\bibliographystyle{JHEP.bst}
\bibliography{LATTICE2021_hybridpotentials}

\end{document}